\documentclass[a4paper]{jpconf}
\usepackage{graphicx}
\begin{document}
\title{Sterile Neutrinos and IceCube}
\author{F.~Halzen$^1$ for the  IceCube Collaboration}

\address{$^1$ University of Wisconsin-Madison, IceCube Research Center, Madison, Wisconsin, USA}

\ead{halzen@icecube.wisc.edu, whuelsnitz@umdgrb.umd.edu, elisa.resconi@tum.de}

\begin{abstract}
Although the framework for oscillations of the three neutrino flavors in the Standard Model has been convincingly established, indications persist that it may be incomplete. Challenges are coming from 
 the  LSND and MiniBooNe short-baseline experiments, from the neutrino sources
 used in the Gallex and Sage solar neutrino experiments and, more recently,
 from an a-posteriori analysis of reactor neutrino experiments. 
 One way to accommodate the reported
 ``anomalies", if real, is to introduce one or more sterile neutrinos in the mass range $\delta\!m^2 \sim 1 eV^2$. TeV atmospheric neutrinos propagating through the Earth undergo resonant oscillations in the presence of sterile neutrinos; a clear signature in a neutrino telescope like IceCube is the the change in shape of the zenith-energy distribution of the atmospheric neutrinos.
 
IceCube detects more than 100,000 atmospheric neutrinos per year. Statistics do not limit such a measurement, but the uncertainties in modeling the expectations of the conventional 3-flavor scenario, including the systematics of the detector, do. We review the status and future perspectives of understanding the zenith and energy response of IceCube in the TeV energy range.
\end{abstract}

\section{Sterile Neutrinos and IceCube}

Neutrino experiments offer opportunities for new discoveries and, occasionally, total surprises. Examples of new physics include sterile neutrinos and additional degrees of freedom in the energy density of the Universe. Even though it is premature to motivate future facilities on the basis of present indications (which include some hints from short-baseline experiments \cite{AguilarArevalo:2010wv} and reactor data \cite{Mueller:2011nm}), recent developments underscore the possibility of unexpected discoveries, supporting the construction of neutrino facilities with the widest science reach. IceCube is such a facility \cite{Halzen:2010yj}. IceCube measures the flux of atmospheric muons and neutrinos with high statistics in a high-energy range that has not been previously explored. The procedure is simply to compare data with expectations based on the extrapolation of lower-energy data. Any deviation observed is an opportunity for discovery in astrophysics or the physics of the neutrinos themselves.

The potential of IceCube to detect eV-mass sterile neutrinos has been recognized for some time \cite{smirnov,Nunokawa:2003ep,Choubey:2007ji}. The problem has been revisited \cite{Kopp:2011qd,Razzaque:2011ab,Barger:2011rc} with the measurement of the atmospheric neutrino flux in the energy range 100\,GeV--400\,TeV with unprecedented statistics, taken when IceCube was half complete \cite{ic_fwd}.  The signature for sterile neutrinos is the disappearance of $\nu_{\mu} (\bar\nu_{\mu})$ resulting from mixing with a new species during propagation between production and detection. For $\delta\!m^2 \sim 1 eV^2$, resonant oscillations are expected in IceCube for TeV atmospheric neutrinos at characteristic zenith angles. An additional signature is the appearance of an excess of $\nu_{e} (\bar\nu_{e})$, with a particular energy and zenith dependence, due to a sterile neutrino acting as an intermediary between muon and electron neutrinos.

However, before addressing the question of whether the data support evidence for a sterile neutrino,  we are investigating the consistency between the data and expectation and the correctness of the model implemented in simulation. Given the high statistics, the focus is on the systematics of the experiment. The systematic issues are being studied with renewed emphasis; up to now, our priority has been the search for neutrino sources beyond the atmosphere, analyses where the simulation of the background is done using the data themselves. We discuss the status of this effort that will also result in improved sensitivity to, for instance, diffuse limits on cosmic neutrinos.

\section{Atmospheric Neutrinos}
\subsection{Zenith Distribution}
\begin{figure}
\begin{center}
\includegraphics[width=1.0\textwidth]{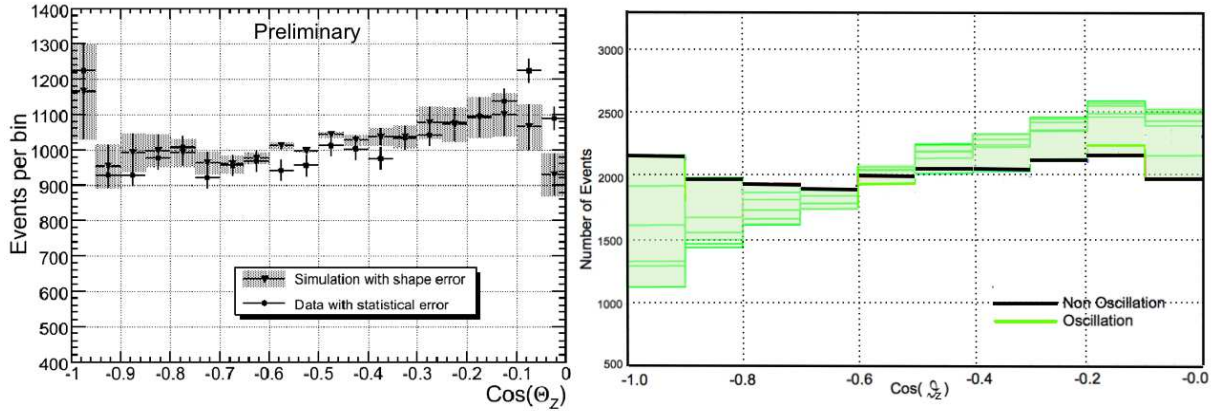}
\caption{Zenith distribution cos($\theta_Z$) of atmospheric neutrinos from \cite{ic_unf}. \textit{On the left}:  Simulation has been normalized to the data.  Error bars for data are statistical only.  Error bars for simulation indicate the uncertainty in the shape of the predicted distribution (normalization uncertainty has been removed).  These error bars do not include certain simulation errors discussed in the text. \textit{On the right}:  Illustration of the impact of adding a sterile neutrino into the conventional 3-flavor scheme on the cos($\theta_Z$) distribution.  For this particular model (green lines), $\Delta m_{41}^2=0.4$ $\mathrm{eV}^2$, $\mathrm{sin}^2\theta_{24}=0.1$, and $\mathrm{sin}^2\theta_{34}$ is varied from 0 to 0.5 in increments of 0.1.  Green and black lines include standard atmospheric neutrino oscillations and detector acceptance.  (credit: A. Esmaili, Universidade Estadual de Campinas, Brazil). }
\label{cos_zen}
\end{center}
\end{figure}

Figure~\ref{cos_zen} compares the observed to the predicted distribution of atmospheric $\nu_\mu$ (plus $\bar{\nu}_{\mu}$) from data taken while IceCube operated in a partially completed 40-string configuration.  Since the uncertainty in the normalization is large (see Ref. 11), it is more interesting to examine the shape of the zenith distribution.  In Fig.~\ref{cos_zen}, the predicted distribution has been normalized to the event rate in data.  This plot is similar to Fig.~19 of Ref. 11, except shape uncertainties are included for the predicted flux.  These shape uncertainties include contributions from uncertainty in the ratio of pions to kaons produced by the cosmic ray flux, uncertainties in the spectral slope of the cosmic ray flux, and uncertainties in the simulation of DOM sensitivity and ice properties.  See \cite{ic_fwd, ic_unf} for further discussion of systematic uncertainties associated with atmospheric neutrino measurements with the 40-string IceCube detector.  Some of these uncertainties will be reduced as the detector and irreducible systematic uncertainties are better understood.  It is important to note that uncertainties due to problems with simulation for the 40-string detector, recently discovered and discussed below, are not included in these error bars.

Several possibilities have been explored to resolve the apparent mismatch between data and simulation in the near horizontal region, including the candidates discussed in \cite{ic_unf, ic_ps}: regional and seasonal variations in atmospheric neutrino production, simulation of ice properties and photon propagation in the ice, and inadequate simulation of atmospheric muon backgrounds including various composition models of cosmic rays.  

While these effects have some impact on the zenith distribution, we have found that an important change arose from improvements in the modeling of the rock layer below the detector and the ice/rock boundary, as well as the assignment of event weights according to model simulations.   These event weights account for the probability of a simulated neutrino to survive propagation through the Earth and interact in or near the detector.  Preliminary testing of improved simulation for the 59-string detector configuration indicates that these inadequacies in the modeling of the detector were a major contributor to the shape disagreement between the 40-string simulation and data.  Atmospheric neutrino analyses and searches for a diffuse flux of astrophysical neutrinos, with data taken by the 59-string detector, are currently underway and are using this improved simulation.

\subsection{Energy Spectrum}
\begin{figure}
\begin{center}
\includegraphics[width=0.5\textwidth]{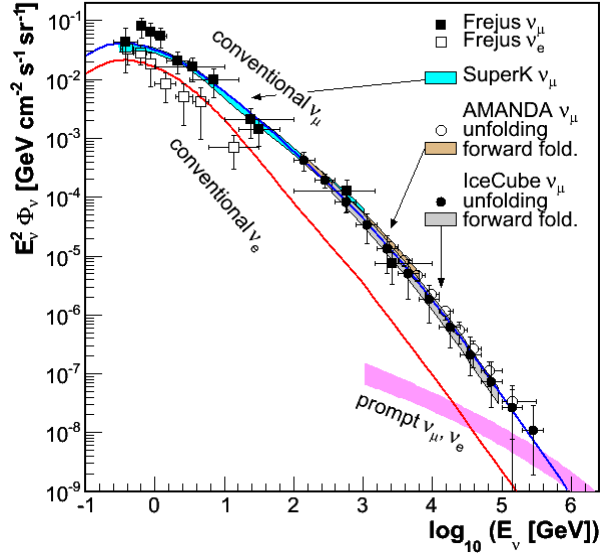}
\caption{Measurements of the atmospheric neutrino energy spectrum; the Fr\'{e}jus results \cite{frejus}, SuperK \cite{superk}, AMANDA forward folding analysis \cite{amanda_fwd} and unfolding analysis \cite{amanda_unf}, IceCube (40 strings) forward folding analysis \cite{ic_fwd} and unfolding analysis \cite{ic_unf}.  All measurements include the sum of neutrinos and antineutrinos.  The expectations for conventional $\nu_\mu$ and $\nu_e$ flux are from \cite{barr}.  The prompt flux is from \cite{sarcevic}.}
\label{compare}
\end{center}
\end{figure}

Figure~\ref{compare} compares the results of atmospheric neutrino measurements from about a GeV to 400~TeV.  In IceCube's spectrum unfolding measurement of \cite{ic_unf}, events in the $90^ \circ $ to $97^ \circ $ zenith region were not used.  However, in Fig.~\ref{compare}, this result has been scaled to the flux for the entire hemisphere, to allow a more direct comparison to the other results.  IceCube's forward folding result from Ref. 10 is also shown.  Both of these results are affected by the simulation issues discussed above.  New analyses using improved simulation are underway using data taken while IceCube operated in the 59-string configuration.

\section{Conclusion}
IceCube has been stably operating with a completed detector since May 2011. Atmospheric neutrinos are collected at the rate of \textit{O}($10^5$) per year.  Our ability to constrain a variety of sterile neutrino models is only limited by systematic uncertainties.  The calibration of the final detector will be superior to that of configurations operated during construction. A new generation of analysis tools are resulting in improved effective area, angular and energy resolution.

\subsection{Acknowledgments}
We acknowledge support from the following agencies: U.S. National Science Foundation, Office of Polar Programs; U.S. National Science Foundation, Physics Division; U. of Wisconsin Alumni Research Foundation; U.S. Department of Energy; U.S. National Energy Research Scientific Computing Center (NERSC); the LONI grid at UCLA; Swedish Research Council; K. \& A. Wallenberg Foundation, Sweden; German Ministry for Education and Research; Deutsche Forschungsgemeinschaft; Fund for Scientific Research, Belgium; IWT-Flanders; Belgian Federal Science Policy Office (BELSPO); the Netherlands Organisation for Scientific Research;  and the Swiss National Science Foundation (SNF).

\end{document}